\documentclass{article}

\input{tcilatex}

\begin{document}

\textbf{MANIPULATING CONSCIOUSNESS}

\bigskip

E. A. Novikov

\bigskip

Institute for Nonlinear Science, University of California - San Diego, La
Jolla, CA 92093 - 0402

\bigskip

\bigskip

Manipulation of the effects of consciousness by external influence on the
human brain is considered in the context of the nonlinear dynamical modeling
of interaction between automatic and conscious processes.

\bigskip

\bigskip

In previous papers [1,2] an approach to nonlinear dynamical modeling of
interaction between automatic (A) and conscious (C) processes in the brain
was presented. The idea is to use quaternion field with real and imaginary
components representing A - and C - processes. The subjective C -
experiences were divided into three major groups: sensations (S), emotions
(E) and reflections (R). Note, that subjective S should be distinguished
from the automatic sensory input into the neuron system of the brain. The A
- C interaction is due to the nonlinearity of the system. This approach was
illustrated on the nonlinear equation for the current density in the cortex.
The nonlinearity is determined by the sigmoidal firing rate of neurons.
Perspective for testing of this approach were also indicated as well as some
more general approaches [1,2].

For the purpose of medical and other possible applications it is interesting
to include an external electromagnetic (EM) influence in this modeling. In a
laboratory setting a specially equipped helmet can produce designed
nonhomogeneous or homogeneous excitations in the brain. On another hand,
suppose we want to pacify a group of terrorists (!) by using a strong EM \
radiation with the wavelength much larger than the size of their brains. In
this case the excitation will be approximately homogeneous. We start with
the homogeneous case which is more simple mathematically and gives some
insight into general situation.

The model equation for the average (spatially uniform) current density $%
\alpha (t)$ perpendicular to the cortical surface has the form [1,2]:

\begin{equation}
\frac{\partial \alpha }{\partial t}+k\alpha =\func{Re}\{f(\alpha +\sigma
+i_{p}\psi _{p})\}+\varphi  \tag{(1)}
\end{equation}%
Here $k$ is the relaxation coefficient, $\sigma (t)$ is the average sensory
input, $f$ represents the sigmoidal firing rate of neurons [for example, $%
f(\alpha )=\tanh (\alpha )$], components $\psi _{p}$ represent the indicated
above (S, E, R) - effects and summation is assumed on repeated subscripts
from 1 to 3. The quaternion imaginary units $i_{p}$ satisfy conditions:

\begin{equation}
i_{p}i_{q}=\varepsilon _{pqr}i_{r}-\delta _{pq}  \tag{(2)}
\end{equation}%
where $\varepsilon _{pqr}$ is the unit antisymmetric tensor and $\delta
_{pq} $ is the unit tensor. Formula (2) is a compact form of conditions: $%
i_{1}^{2}=i_{2}^{2}=i_{3}^{2}=-1,$ $i_{1}i_{2}=-i_{2}i_{1}=i_{3},$ $%
i_{2}i_{3}=-i_{3}i_{2}=i_{1},$ $i_{3}i_{1}=-i_{1}i_{3}=i_{2}.$ Equation (1)
is obtained by using the quaternion $q=\alpha +i_{p}\psi _{p}$ instead of $%
\alpha $ in order to describe the A - C interaction. The additional term $%
\varphi $ in (1) represents the external EM excitation. Equation (1) is the
real part of the equation for the quaternion [1,2]:

\begin{equation}
\frac{\partial q}{\partial t}+kq=f(q+\sigma )+\phi  \tag{(1a)}
\end{equation}

For $\psi _{p}$ from (1a) we have equations:

\begin{equation}
\frac{\partial \psi _{p}}{\partial t}+k\psi _{p}=\func{Im}_{p}\{f(\alpha
+\sigma +i_{q}\psi _{q})\},\text{ }p=1,2,3  \tag{(3)}
\end{equation}%
where $\func{Im}_{p}\{f\}=-\func{Re}\{fi_{p}\}$. Note, that so-called
extra-sensory effects (if they exist) can be included in this approach by
assuming that $\sigma $ is a quaternion: $\sigma \Longrightarrow \sigma
+i_{p}s_{p}$, this will produce shift $\psi _{p}\Longrightarrow \psi
_{p}+s_{p}$ in the nonlinear terms in (1), (3) and below in (5), (6).

Let us consider typical $f(\alpha )=\tanh (\alpha )$. Simple algebra gives
[2]:

\begin{equation}
\tanh (q)=\frac{\sinh (2\alpha )+j\sin (2\psi )}{\cosh (2\alpha )+\cos
(2\psi )},\text{ }\psi ^{2}\equiv \psi _{p}^{2},\text{ }j\equiv i_{p}\psi
_{p}\psi ^{-1},\text{ }j^{2}=-1  \tag{(4)}
\end{equation}%
Using (4) with shift $\alpha \Longrightarrow \alpha +\sigma $, we rewrite
(1) and (3) explicitly:%
\begin{equation}
\frac{\partial \alpha }{\partial t}+k\alpha =\frac{\sinh [2(\alpha +\sigma )]%
}{\cosh [2(\alpha +\sigma )]+\cos (2\psi )}+\phi  \tag{(5)}
\end{equation}

\begin{equation}
\frac{\partial \psi _{p}}{\partial t}+k\psi _{p}=\frac{\psi _{p}\psi
^{-1}\sin (2\psi )}{\cosh [2(\alpha +\sigma )]+\cos (2\psi )},\text{ }p=1,2,3
\tag{(6)}
\end{equation}

Some general conclusions can be made without solving these equations.
Firstly, if $\psi _{p}(0)=0$ than $\psi _{p}(t)\equiv 0$ (unless $s_{p}\neq
0 $). Secondly, if $\psi _{p}(0)\neq 0$, than evolution $\psi _{p}(t)$ can
be manipulated by using sensory input $\sigma (t)$ and EM excitation $\phi
(t)$. Thirdly, the nonlinearity of the system suggests that the efficiency
of such manipulation depends not only on the amplitudes of $\sigma (t)$ and $%
\phi (t)$ but also on the shape of these functions (spectral content).

For the case of spatially nonuniform $\alpha (t,\mathbf{x}),\psi _{p}(t,%
\mathbf{x}),\sigma (t,\mathbf{x})$ and $\phi (t,\mathbf{x})$ we can use more
general equations, which include typical propagation velocity of signals in
the neuron system of the cortex $v$. Time differentiation of (1a), simple
algebra and addition a term with the two-dimensional spatial Laplacian $%
\Delta $ gives [1,2]:

\begin{equation}
\frac{\partial ^{2}q}{\partial t^{2}}+(k+m)\frac{\partial q}{\partial t}%
+(km-v^{2}\Delta )q=(m+\frac{\partial }{\partial t})f(q+\sigma )+\frac{%
\partial \phi }{\partial t}  \tag{(7)}
\end{equation}%
where $m$ is an arbitrary parameter (see below). Real and imaginary
projections of (7) give equations for $\alpha $ and $\psi _{p}$, which are
generalizations of (1) and (3). If we put $\psi _{p}=0$ and $\phi =0$, than
equation for $\alpha $ will be similar in spirit to equations used for
interpretation of EEG and MEG spatial patterns ( see recent paper [3] and
references therein). In this context we have parameters: $k\sim m\sim v/l$,
where $l$ is the connectivity scale. For $f(\alpha )=\tanh (\alpha )$ the
nonlinear term $f(q+\sigma )$ in (7) has the same projections as in (5) and
(6).

The obtained in this letter equations can be used for numerical experiments
and for comparison with corresponding laboratory experiments.

\bigskip

\textbf{References}

\bigskip

[1] E. A. Novikov, Towards modeling of consciousness, arXiv:nlin.PS/0309043
(2003)

[2] E. A. Novikov, Quaternion dynamics of the brain, arXiv:nlin.PS/0311047
(2003)

[3] V. K. Jirsa, K. J. Jantzen, A. Fuchs, and J. A. Kelso, Spatiotemporal
forward solution

of the EEG and MEG using network modeling, IEEE Trans. Med. Imaging, \textbf{%
21}(5), 497 (2002)

\end{document}